\documentclass[twocolumn,english,prl,aps,showpacs]{revtex4}
\usepackage[T1]{fontenc}
\usepackage[latin9]{inputenc}
\usepackage{amsmath}
\usepackage{amssymb}
\usepackage{graphicx}
\usepackage{esint}

\makeatletter
\@ifundefined{textcolor}{}
{%
 \definecolor{BLACK}{gray}{0}
 \definecolor{WHITE}{gray}{1}
 \definecolor{RED}{rgb}{1,0,0}
 \definecolor{GREEN}{rgb}{0,1,0}
 \definecolor{BLUE}{rgb}{0,0,1}
 \definecolor{CYAN}{cmyk}{1,0,0,0}
 \definecolor{MAGENTA}{cmyk}{0,1,0,0}
 \definecolor{YELLOW}{cmyk}{0,0,1,0}
 }

\makeatother

\usepackage{babel}
\begin{document}

\title{Phase matching condition for enhancement of phase sensitivity in
quantum metrology}

\author{Jing Liu, Xiaoxing Jing, Xiaoguang Wang}

\email{xgwang@zimp.zju.edu.cn}

\selectlanguage{english}

\affiliation{Zhejiang Institute of Modern Physics, Department of Physics, Zhejiang
University, Hangzhou 310027, China}
\begin{abstract}
We find a phase matching condition for enhancement of sensitivity
in a Mach-Zehnder interferometer illuminated by an arbitrary state
in one input port and an odd(even) state in the other port. Under
this condition, the Fisher information becomes maximal with respect to the relative phase of two modes and the phase sensitivity is enhanced. For the case with photon losses, we further find that the
phase matching condition keeps unchanged with a coherent state and
a coherent superposition state as the input states.
\end{abstract}


\maketitle
\emph{Introduction}.--With the development of quantum information
theory~\cite{1,1-1,1-1.1,1-2,1-3} and quantum technology~\cite{2,2-1,2-2}, quantum
metrology is becoming more and more practical nowadays. Improving
the precision of a parameter is always the basic theme in quantum
metrology. The fundamental procedure of it is the parametrization
process. There are three mainly methods to perform this procedure:
(i) unitary parametrization; (ii) channel parametrization; (iii) accelerating
parametrization. Unitary parametrization is the most useful and well
studied method because it is widely applied in the phase estimation,
the main task in quantum metrology. The precision measurement of gravity,
temperature, week magnetic strength and many other parameters can
be classified in the category of phase estimation. Channel parametrization
is also widely studied in the channel estimation~\cite{Fujiwara,channel_est}
for many years. With the help of the quantum technology, it may be
possible to proceed the parametrization in a accelerating way~\cite{relative},
in which the relativistic effects cannot be neglected any more. In
this paper, we mainly focus on the category of phase estimation and
study how to effectively enhance the phase sensitivity.

In 1981, Caves~\cite{Caves} found out that, for phase estimation,
taking a high intensity coherent state and a low intensity squeezed
vacuum state as the input states of a Mach-Zehnder interferometer,
the precision can beat the shot-noise limit(quantum standard limit),
i.e., $1/\sqrt{N}$, where $N$ is the total photon number in both
modes. Since then, many protocols have been proposed to fulfill the
similar job, such as NOON state, entangled coherent state~\cite{Joo},
two-mode squeezed state~\cite{Anisimov}, number squeezed states~\cite{Pezze2013}
and so on. With high intensity, some of these states can even theoretically
achieve or surpass the Heisenberg limit, i.e., $1/N$. In the pioneer
work of Caves~\cite{Caves}, to enhance the precision, the phases
of the two input states need to satisfy a relation. This can be considered
as a kind of phase matching condition (PMC). Thus, it is reasonable to study
if there is a more general PMC for more general states to to enhance the phase
sensitivity. This is the major motivation of this paper. To depict
the precision of a parameter $\theta$, quantum Fisher information(QFI)
is an available useful concept because it describes the lower bound on the
variance of the estimator $\hat{\theta}$ due to the Cram\'{e}r-Rao
theorem: $\mathrm{Var}(\hat{\theta})\geq1/(\nu F)$,~\cite{Helstrom,Holevo}
where $\mathrm{Var}(\cdot)$ is the variance, $\nu$ is the number
of repeated experiments and $F$ is the QFI.

\begin{figure}[bp]
 \includegraphics[width=7.5cm]{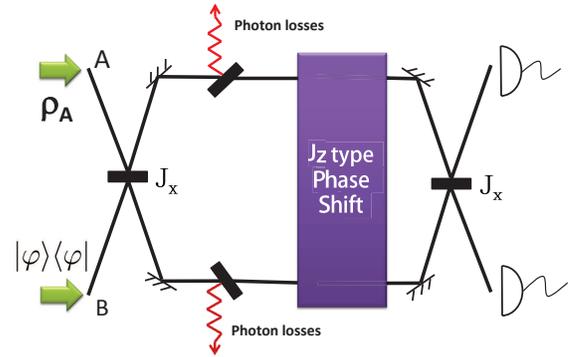} \caption{\label{fig:Sketch}(Color online) Sketch of the Mach-Zehnder interferometer.
The transformation of the beam splitters are described by $\exp(\pm i\pi J_{x}/2)$,
and that of the phase shift is described by $\exp(i\theta J_{z})$.
The input states in port A and B are an arbitrary state and an even(odd)
state, respectively.}
\end{figure}

In this paper, we discuss a general scenario of a Mach-Zehnder interferometer.
In this interferometer, one of the input port is an \emph{arbitrary}
state and the other one is an \emph{even}(odd) state. This scenario covers
many important cases including the famous protocol of Caves~\cite{Caves},
and a recent one proposed by Pezz\'{e} and Smerzi~\cite{Pezze2013}.
We give an analytic expression of the QFI and identify a PMC to optimize the parameter
precision.
Under this condition, the QFI is only determined
the the average photon numbers of the two modes and the corresponding
expectation values of the square of annihilation operators. We then give two examples
of our scenario. Further, for the case that photon losses occur in
both arms with the same transmission coefficients, and the input state
is a product of a coherent state and a coherent superposition state, the analytic
expression of QFI is provided. Based on this expression, we prove that the
PMC keeps unchanged for any transmission coefficients.

\emph{Mach-Zehnder interferometer}.--Mach-Zehnder(MZ) interferometer
is a well known optical device in quantum metrology which is constructed
with two beam splitters and one or two phase shifts. The interferometer
we consider here is constructed with two 50:50 beam splitters and
two phase shifts in both arms, as shown in Fig.~\ref{fig:Sketch}.
The well used 50:50 beam splitter can be described by~\cite{Campos,Sanders,Yurke}
$B_{x}=\exp(-i\frac{\pi}{2}J_{x})$, where $J_{x}=\frac{1}{2}(a^{\dagger}b+b^{\dagger}a)$
is one of the operators in the Schwinger representation of bosons.
The others are $J_{y}=\frac{1}{2i}(a^{\dagger}b-b^{\dagger}a)$ and
$J_{z}=\frac{1}{2}\left(a^{\dagger}a-b^{\dagger}b\right)$. Here $a$,
$b$ are annihilation operators for ports A and B, respectively. Operators $J_{x}$, $J_{y}$,
$J_{z}$ satisfy the commutation: $\left[J_{i},J_{j}\right]=i\epsilon_{ijk}J_{k}$,
with $\epsilon_{ijk}$ the so-called Levi-Civita symbol. Usually,
a phase shift can be expressed by a unitary transformation $\exp[i\theta\hat{N}_{\mathrm{A(B)}}]$,
where $\hat{N}_{\mathrm{A}}=a^{\dagger}a$, $\hat{N}_{\mathrm{B}}=b^{\dagger}b$
and $\theta$ is an unknown parameter. In our scenario, denoting $\theta$
as the relative phase between the two arms, one can describe the transformation
of the total phase shift by the operator $P_{z}=\exp(i\theta J_{z})$.
With above devices we can construct a well used MZ interferometer,
the transformation of which can be written as $U_{\mathrm{mz}}=B_{x}P_{z}B_{x}^\dagger$.
Through some algebra, one can find that this transformation can be
simplified as
\begin{eqnarray}
U_{\mathrm{MZ}} & = & \exp\left(-i\theta J_{y}\right),
\end{eqnarray}
which is a rotation along $y$ direction.

\emph{Phase matching condition for QFI}.--Quantum Fisher information(QFI)
is a central concept in quantum metrology, and it is defined as~\cite{Helstrom,Holevo}
$F:=\mathrm{Tr}\left(\rho L^{2}\right)$, where $L$ is the so-called
symmetric logarithmic derivative determined by $\partial_{\theta}\rho_{\theta}=\left(\rho_{\theta}L+L\rho_{\theta}\right)/2.$
We consider a separable input state $\rho_{\mathrm{in}}=\rho_{\mathrm{A}}\otimes\rho_{\mathrm{B}}$.
Here $\rho_{\mathrm{A}}$ is an arbitrary state and $\rho_{\mathrm{B}}=|\varphi\rangle\langle\varphi|$
is an even(odd) state. Utilizing the spectral decomposition $\rho_{\mathrm{A}}=\sum_{j=1}^Mp_{j}|\psi_{j}\rangle\langle\psi_{j}|$,
the quantum Fisher information can be written as~\cite{Liu,Helstrom,Holevo}
\begin{equation}
F=\sum_{j=1}^{M}4p_{j}\langle\phi_{j}|J_{y}^{2}|\phi_{j}\rangle-\sum_{j,j^{\prime}=1}^{M}\frac{8p_{j}p_{j^{\prime}}}{p_{j}+p_{j^{\prime}}}|\langle\phi_{j}|J_{y}|\phi_{j^{\prime}}\rangle|^{2},
\end{equation}
where $M$ is the dimension of the support of $\rho_{\mathrm{A}}$
and $|\phi_{j}\rangle=|\psi_{j}\rangle\otimes|\varphi\rangle$ is
an eigenstate of $\rho_{\mathrm{in}}$ with eigenvalue $p_j$.

As the input state in port B is an even(odd) state in our scenario,
it satisfies $\langle\varphi|b|\varphi\rangle=0$. Then after some
algebra, the QFI can be expressed by
\begin{eqnarray}
F & = & 2\bar{n}_{\mathrm{A}}\bar{n}_{\mathrm{B}}+\bar{n}_{\mathrm{A}}+\bar{n}_{\mathrm{B}}-2\mathrm{Re}\left(\langle a^{\dagger2}\rangle\langle b^{2}\rangle\right),\label{eq:F_y}
\end{eqnarray}
 where $\bar{n}_{\mathrm{A}}=\mathrm{Tr}(\rho_{\mathrm{A}}a^{\dagger}a)$
and $\bar{n}_{\mathrm{B}}=\langle\varphi|b^{\dagger}b|\varphi\rangle$ are average photon numbers for mode $a$ and $b$, respectively.
To optimize this Fisher information, the input states need to satisfy
the following PMC:
\begin{equation}
|\mathrm{Arg}(\langle a^{2}\rangle)-\mathrm{Arg}(\langle b^{2}\rangle)|=\pi.\label{eq:PMC}
\end{equation}
Under this condition, the QFI becomes
\begin{equation}
F_{\mathrm{m}}=2\bar{n}_{\mathrm{A}}\bar{n}_{\mathrm{B}}+\bar{n}_{\mathrm{A}}+\bar{n}_{\mathrm{B}}+2|\langle a^{2}\rangle||\langle b^{2}\rangle|.\label{eq:FI}
\end{equation}
One key feature of the above result is that the QFI is only determined by the mean photon numbers and the expectations of $a^2$ and $b^2$. We emphasize that
one input state is arbitrary.

For the case that $|\langle a^{2}\rangle||\langle b^{2}\rangle|=0$,
the phases of the input states can be chosen arbitrarily. For example,
if $\rho_{\mathrm{B}}$ is a Fock state, i.e., $\rho_{\mathrm{B}}=|n\rangle\langle n|$,
then the QFI reduces to $F_{\mathrm{m}}=2\bar{n}_{\mathrm{A}}\bar{n}_{\mathrm{B}}+\bar{n}_{\mathrm{A}}+\bar{n}_{\mathrm{B}}$,
which is the result in Ref.~\cite{Pezze2013} and has been discussed
in detail. In the following, we give two examples of the PMC.

\textbf{Example 1}.--In this example, we choose $\rho_{\mathrm{A}}$
be a coherent state $|\beta\rangle\langle\beta|$ and $\rho_{\mathrm{B}}$ a coherent superposition state $|{\alpha}\rangle_+\langle\alpha|$, where $|{\alpha}\rangle_+=N_{\alpha}(|\alpha\rangle+|-\alpha\rangle)$ with
$|\alpha\rangle$ also a coherent state and $N_{\alpha}^{2}=1/(2+2e^{-2|\alpha|^{2}})$.
It is easy to find that  $_+\langle{\alpha}|b|{\alpha}\rangle_+=0$. Here, we denote $\alpha=|\alpha|\exp(i\Phi_{\alpha})$,
$\beta=|\beta|\exp(i\Phi_{\beta})$, then from Eq.~(\ref{eq:PMC}), the phase matching condition
can be specifically written as
\begin{equation}
|\Phi_{\alpha}-\Phi_{\beta}|=\frac{\pi}{2}.
\end{equation}
Under this condition, the QFI reduces to
\begin{equation}
F_{\mathrm{m}}=2\bar{n}_{\mathrm{A}}\bar{n}_{\mathrm{B}}+\bar{n}_{\mathrm{A}}+\bar{n}_{\mathrm{B}}+2\bar{n}_{\mathrm{A}}|\alpha|^{2},\label{eq:ex_1}
\end{equation}
where $\bar{n}_{\mathrm{A}}=|\beta|^{2}$ and $\bar{n}_{\mathrm{B}}=|\alpha|^{2}\tanh|\alpha|^{2}$.
As $\tanh|\alpha|^{2}$ is a monotonic function and very close to
$1$ for $|\alpha|\geq2$, then for
most value of $|\alpha|$, $\bar{n}_{\mathrm{B}}$ is equal to $|\alpha|^{2}$,
and the QFI reduces to $F_{\mathrm{m}}=4\bar{n}_{\mathrm{A}}\bar{n}_{\mathrm{B}}+\bar{n}_{\mathrm{A}}+\bar{n}_{\mathrm{B}}$.
It is not difficult to obtain that $F_{\mathrm{m}}\leq N^{2}+N$,
where $N=\bar{n}_{\mathrm{A}}+\bar{n}_{\mathrm{B}}$ is the total photon
number. The equality can be achieved when $\bar{n}_{\mathrm{A}}=\bar{n}_{\mathrm{B}}$.
For the case that total photon number $N$ is fixed, this bound is
the optimal value of the QFI. Considering the PMC, the optimal value can be achieved when $\beta=\pm i\alpha$.
From the optimal value of $F_{\mathrm{m}}=N^{2}+N\geq N^{2}$,
one can find that, with high intensity, the pair of the coherent state
$|\pm i\alpha\rangle\langle\pm i\alpha|$ and the coherent superposition
state $|{\alpha}\rangle_+$
can surpass the Heisenberg limit. Ref.~\cite{Joo} considered the
same input states in our scenario but with only one phase shift in
one arm. In their case, utilizing the similar analysis above, it is easy
to find that the input states $|\alpha\rangle$ and
$|{\alpha}\rangle_+$ is the
optimal choice and can reach the maximum value of QFI.

\begin{figure}
\includegraphics[width=7cm]{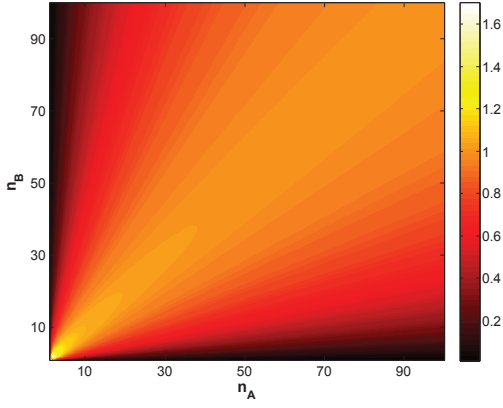} \caption{\label{fig:squeeze_vacuum}(Color online) The variation of $F_{\mathrm{m}}/N^2$ with the average particle number $\bar{n}_{\mathrm{A}}$
and $\bar{n}_{\mathrm{B}}$. The input states here are a coherent
state $|\beta\rangle\langle\beta|$ in port A and a squeezed vacuum
state $|\xi\rangle\langle\xi|$ in port B.}
\end{figure}

\textbf{Example 2}.--Another well known even state is the squeezed
vacuum state, which is defined as~\cite{Scully} $|\xi\rangle=S(\xi)|0\rangle$,
where the squeezing operator reads $S(\xi)=\exp(\frac{1}{2}\xi^{*}b^{2}-\frac{1}{2}\xi b^{\dagger^{2}})$
with $\xi=|\xi|\exp(i\Phi_{\xi})$. For convenience, we still choose
the input state in port A as a coherent state $|\beta\rangle\langle\beta|$.
In this example, the PMC is
\begin{equation}
2\Phi_{\beta}-\Phi_{\xi}=0.
\end{equation}
Under this condition, the QFI can be expressed as
\begin{equation}
F_{\mathrm{m}}=2\bar{n}_{\mathrm{A}}\bar{n}_{\mathrm{B}}+\bar{n}_{\mathrm{A}}+\bar{n}_{\mathrm{B}}+2\bar{n}_{\mathrm{A}}\sqrt{\bar{n}_{\mathrm{B}}^{2}+\bar{n}_{\mathrm{B}}},
\end{equation}
where $\bar{n}_{\mathrm{A}}=|\beta|^{2}$ and $\bar{n}_{\mathrm{B}}=\sinh^{2}|\xi|$.
This equation is equivalent to the corresponding equation in Ref.~\cite{Jarzyna}.
In this expression of QFI, it is only related to the average photon
numbers in both ports: $\bar{n}_{\mathrm{A}}$ and $\bar{n}_{\mathrm{B}}$.
Figure~\ref{fig:squeeze_vacuum} shows the variation of $F_{\mathrm{m}}/N^2$
with the change of $\bar{n}_{\mathrm{A}}$, $\bar{n}_{\mathrm{B}}$.
From this plot one can find that the optimal value of quantum Fisher information
for a fixed $N$ is obtained near the $n_{\mathrm{A}}=n_{\mathrm{B}}$
line, especially for a large $N$. This is because when $\bar{n}_{\mathrm{B}}$
is large, for fixed $N$, $F_{\mathrm{m}}\simeq4\bar{n}_{\mathrm{A}}\bar{n}_{\mathrm{B}}+\bar{n}_{\mathrm{A}}+\bar{n}_{\mathrm{B}}\leq N^{2}+N$ and the optimal value can be achieved when $\bar{n}_{\mathrm{A}}=\bar{n}_{\mathrm{B}}$.
Adding the PMC, the optimal choice for a large
$N$ is that $\beta=\exp(i\Phi_{\xi}/2)\sinh|\xi|$. Also, from this figure one can see that with the increase of $N$, the range that $F_{\mathrm{m}}>N^2$ is increasing, which indicates that high intensity input state is good for the enhancement of the phase sensitivity.

\emph{PMC with unbanlanced beam splitter.}--For a more general beam splitter transformation $\exp(i\tau J_{x})$
with $\tau\in[0,2\pi)$, the total setup of the interferometer can
be described by
\begin{equation}
U_{\mathrm{MZ}}=\exp[i\theta(J_{z}\cos\tau-J_{y}\sin\tau)].
\end{equation}
If we restrict the input state $\rho_{\mathrm{A}}$ as a pure state,
i.e., $|\psi\rangle\langle\psi|$, and $\rho_{\mathrm{B}}$ is still
an even(odd) state, the QFI reads $F=4\mathrm{Var}(J_{z}\cos\tau-J_{y}\sin\tau)$.
Due to the property of even(odd) state, one can find that
\begin{equation}
F=4\cos^{2}\tau\mathrm{Var}(J_{z})+4\sin^{2}\tau\mathrm{Var}(J_{y}).
\end{equation}
It is known that for product states
\begin{equation}
\mathrm{Var}(J_{z})=\frac{1}{4}\left[\mathrm{Var}(a^{\dagger}a)
+\mathrm{Var}(b^{\dagger}b)\right],
\end{equation}
and $\mathrm{Var}(a^{\dagger}a)$, $\mathrm{Var}(b^{\dagger}b)$ are
irrelevant to the phase, therefore, the phase matching condition is
decided by the second item $\mathrm{Var}(J_{y})$ and keeps the same
form with Eq.~(\ref{eq:PMC}).

\emph{PMC with photon losses.}--Now we consider
the effects of photon losses on the PMC. The scenario
we use is shown in Fig.~1. Traditionally,
the photon losses can be described by a fictitious beam splitter~\cite{Dobrzanski09,Dorner09,Gkortsilas12,Huver08,Lee09,Xiang-Bin Wang}.
Here we use $B^{T}=\exp[i(2\arccos\sqrt{T})J_{x}]$ to describe this
fictitious beam splitter. $T$ is the so-called transmission coefficient
and we also define $R=1-T$ as the reflection coefficient. When $T=1(R=0)$,
there are no photon losses in the interferometer and when $T=0(R=1)$,
all the photons leak out of the interferometer. This total photon loss scenario can be mapped into a neat scenario, which includes two steps: first the input state goes through a particle loss channel, then the output state imports into a MZ interferometer without losses.

We assume that the leak
in both arms share the same transmission coefficient $T$ and the
input state is separable: $\rho_{\mathrm{A}}\otimes\rho_{\mathrm{B}}$.
Then after the particle loss channel, the reduced density matrix $\rho$ reads
\begin{equation}
\rho=\mathrm{Tr}_{\mathrm{CD}}\left(\Gamma^{\mathrm{D}}\Gamma^{\mathrm{C}}\rho_{\mathrm{A}}\otimes\rho_{\mathrm{B}}\Gamma^{\mathrm{C}\dagger}\Gamma^{\mathrm{D}\dagger}\right),
\end{equation}
where the operator of the particle loss channel can be expressed by
$\Gamma^{\mathrm{C}}=\exp\left[i\sqrt{2}\arccos\sqrt{T}\left(J_{x}^{\mathrm{AC}}+J_{y}^{\mathrm{BC}}\right)\right]$
and $\Gamma^{\mathrm{D}}=\exp\left[i\sqrt{2}\arccos\sqrt{T}\left(J_{x}^{\mathrm{BD}}+J_{y}^{\mathrm{AD}}\right)\right]$.
Here $J_{x}^{\mathrm{AC}}=\frac{1}{2}(a^{\dagger}c+ac^{\dagger})$
and $J_{y}^{\mathrm{BC}}=\frac{1}{2i}(b^{\dagger}c-bc^{\dagger})$
with $c$, $c^{\dagger}$ the annihilation and creation operators
of mode C. So as $J_{x}^{\mathrm{BD}}$ and $J_{y}^{\mathrm{AD}}$.
Next $\rho$ goes through a usual MZ interferometer, which can be described by $\exp(-\theta J_{y})$.
Utilizing this representation, this photon losses scenario can be classified into a usual MZ interferometer scenario with a mixed input state.

Now we choose the initial states as $\rho_{\mathrm{A}}=|i\alpha e^{i\Phi}\rangle\langle i\alpha e^{i\Phi}|$
and $\rho_{\mathrm{B}}=|{\alpha}\rangle_+\langle{\alpha}|$.
Here $\Phi$ is the relative phase and $\Phi\in[0,\pi)$. Based on
the PMC, the optimal value of it is zero for
no loss scenario. The similar scenario has been considered with only
one phase shift in one arm ~\cite{Joo,G R Jin,Jing}. Through some  calculations,
one can obtain the analytic expression of the QFI
\begin{align}
F= & 4T|\alpha|^{2}\!\left[N_{\alpha}^{2}\!+\! T|\alpha|^{2}\!\left(2N_{\alpha}^{2}-1\right)\!\right]\!+\!4T^{2}|\alpha|^{4}\mathcal{G}\cos^{2}\Phi\nonumber \\
 & -16T^{2}N_{\alpha}^{4}|\alpha|^{4}(1-p_{r}^{2})p_{t}^{2}\sin^{2}\Phi,\label{eq:FI_loss}
\end{align}
where $p_{r}=\exp(-2|\alpha|^{2}R)$, $p_{t}=\exp(-2|\alpha|^{2}T)$
and $\mathcal{G}=1-4N_{\alpha}^{4}\left(1-p_{r}^{2}\right)$. From
this equation, one can see that for a fixed $\alpha$ and transmission
coefficient $T$, the maximum value of Eq.~(\ref{eq:FI_loss}) can
always be obtained at $\Phi=0$, which indicates that the phase matching
condition remains \emph{unchanged }in this photon losses case and
is not effected by the transmission coefficient $T$. Under the phase
matching condition, the quantum Fisher information reads
\begin{equation}
F_{\mathrm{m}}=4TN_{\alpha}^{2}|\alpha|^{2}+8T^{2}N_{\alpha}^{2}|\alpha|^{4}\left[1-2N_{\alpha}^{2}\left(1-p_{r}^{2}\right)\right].
\end{equation}
Utilizing the input average photon number $\bar{n}_{\mathrm{A}}$
and the input total photon number $N$, it can be rewritten as
\begin{equation}
F_{\mathrm{m}}=TN+2T^{2}N\bar{n}_{\mathrm{A}}\left[1-2N_{\alpha}^{2}\left(1-p_{r}^{2}\right)\right].
\end{equation}
This equation can reduce to Eq.~(\ref{eq:ex_1}) for $T=1(R=0)$.
From this equation, one can find that the photon losses have a negative
influence on the QFI. With the decrease of the transmission coefficient
$T$, the QFI reduces. For a small losses, namely $R$ is very small,
the QFI reduces to
\begin{equation}
F_{\mathrm{m}}=N+2N\bar{n}_{\mathrm{A}}-\left[1+4\bar{n}_{\mathrm{A}}(N+1)\right]NR.
\end{equation}
When $R<R_{c}$, with $R_{c}=2\bar{n}_{\mathrm{A}}/\left[1+4\bar{n}_{\mathrm{A}}(N+1)\right]$,
the QFI can still larger than $N$, which indicates that these input
states can still surpass the shot-noise limit. This device is
loss-tolerant and robust within the region that $R\in[0,R_{c}]$.
For a large $N$, $R_{c}\simeq1/(2N)$.

\emph{Conclusion}.--In summary, we have considered a general scenario of a Mach-Zehnder
interferometer, where the input state in one port of the interferometer
is an arbitrary state and the other one is an even(odd) state. We
have provided an analytic expression of the QFI
and showed a general PMC under which the phase sensitivity can be enhanced. We give two explicit
examples with one port in a coherent state and another in a even coherent state or the
squeezed vacuum state. The PMC plays an important role to improve the sensitivity in these cases.

For the unbalanced beam splitters, we find the PMC is unchanged with one port in a arbitrary
pure stare and another in an even(odd) state.
We also considered the example of coherent superposition state with
photon losses and  find that the
PMC keeps unchanged and is not affected by the
transmission coefficients. Besides, for a small loss, we show that
this setup is robust and loss-tolerant. The present work sheds new light on
the problem of how to enhance phase sensitivity in quantum metrology.

\emph{Acknowledgments}.--This work was supported by the NFRPC through
Grant No. 2012CB921602, the NSFC through Grants No. 11025527 and No.
10935010.

\end{document}